\documentstyle[aps,multicol,pra,epsfig]{revtex}

\begin{document}

\draft

\title{Soliton Propagation in Chains with Simple 
Nonlocal Defects}
\author{R. Burioni$^1$, D. Cassi$^1$, P. Sodano$^2$, 
A. Trombettoni$^2$, and A. Vezzani$^1$}
\address{$^1$ I.N.F.M. and Dipartimento di Fisica, Universit\`a
di Parma, parco Area delle Scienze 7A Parma, I-43100, Italy}
\address{$^2$ Dipartimento di Fisica and Sezione I.N.F.N.,
Universit\`a di Perugia, Via A. Pascoli Perugia, I-06123, Italy}

\date{\today}
\maketitle

\begin{abstract} 
We study the propagation of solitons 
on complex chains built by inserting finite graphs at two sites 
of an unbranched chain. We compare numerical findings with 
the results of an analytical linear approximation scheme 
describing the interaction of large-fast solitons with non-local 
topological defects on a chain. 
We show that the transmission properties of the solitons strongly depend 
on the structure of the inserted graph, giving a tool to control 
the soliton propagation through the choice of pertinent graphs 
to be attached to the chain.     
\end{abstract}


\begin{multicols}{2}

\section{Introduction}  
 
Recently, much attention has been devoted to the analysis 
of complexity arising in discrete physical systems living on 
networks with non-trivial topologies: remarkable examples are 
by now given by networks of nonlinear waveguides 
\cite{kivshar03}, Bose-Einstein condensates in optical lattices 
\cite{oberthaler03}, Josephson junction networks (JJN) \cite{fazio01} 
and silicon-based photonic crystals \cite{birner01}. In these systems, 
the choice of the network's topology allows 
to engineer new macroscopically coherent quantum states: a remarkable 
example of macroscopic quantum coherence topologically induced by the 
network topology in JJN has been predicted in \cite{burioni00}.

Another relevant area where one should be able to evidence new phenomena 
induced in discrete quantum systems by the topology is provided 
by nonlinear dynamical systems. Nonlinearity already produces 
remarkable phenomena such as soliton propagation \cite{scott99} 
even in very simple geometries. A few steps 
in the study of the interplay between nonlinearity and 
complex topology have been made: 
recently, the effects of uniformity break on soliton propagation 
\cite{christodoulides01,kevrekidis03} 
and localized modes \cite{mcgurn00} have been investigated 
by considering $Y$-junctions  
\cite{christodoulides01,kevrekidis03} (consisting of a  
long chain inserted on a site of a chain yielding 
a star-like geometry) or geometries like  
junctions of two infinite waveguides or the waveguide coupler 
\cite{mcgurn00}. 
In \cite{chaos} it has been considered the
discrete nonlinear Schr\"odinger equation (DNLSE) on a discrete network
obtained by inserting a finite graph at a site of a one dimensional 
unbranched chain and the soliton propagation 
through this finite graph has been studied by numerical and 
analytical tools. It has been showed that 
for sufficiently large and fast solitons 
the soliton momenta for perfect reflection and transmission can be 
analytically related to the energy levels of the 
inserted graph. Such results for the transmission
properties of solitons in inhomogeneous networks 
have been used in \cite{burioni05} to show that it is possible 
to engineer topological filters for the soliton motion on a complex 
chain.

The simple criterion relating the perfect reflection and transmission momenta 
and the energy levels of the inserted graph obtained in \cite{chaos} 
has been obtained in the situation in which the graph is inserted 
at a single site. In the present paper we address the more difficult 
issue of the soliton propagation for the DNLSE on a network obtained by 
inserting a graph on {\em two} sites of an unbranched chain. In this 
respect, the inserted graph has an internal structure and it 
is seen by the soliton like an extended topological defect, since 
during the motion along the unbranched chain the soliton 
wavefunction is modified in more than one site. One therefore expects 
a variety of possible resonances between 
the energies characterizing the soliton propagation and the energy levels 
of the network. We shall focus here only on some particularly simple 
inserted graphs: loops, bubbles and single links attached in two sites 
(see Fig.1). The plan of the paper is the following: 
after introducing in the next Section the DNLSE 
on a graph, we report in Section III our results for the transmission
properties of solitons 
on such networks by resorting to numerical simulations and analytical results 
based on the linear approximation for the analysis 
of the interaction of fast solitons with these topological defects. 
Section IV is devoted to our conclusions.

\section{The DNLSE on a graph}

The DNLSE is a paradigmatic example of a nonlinear wave equation 
extensively studied on regular lattices 
\cite{hennig99,kevrekidis01,ablowitz04}. On a chain it reads
\begin{equation}
\label{DNLS-retta}
i  \frac{\partial \psi_n}{\partial t} = - \frac{1}{2} (\psi_{n+1} 
+ \psi_{n-1}) + \Lambda \mid \psi_n \mid ^2 \psi_n 
\end{equation}
where $n=\cdots,-1,0,1\cdots$ is an integer index denoting 
the site position and $\Lambda$ is the coefficient of the 
nonlinear term. 
The normalization condition is 
$\sum_n \mid \psi_n \mid ^2=1$. 
It is well known that the DNLSE on a homogeneous chain 
is not integrable; however, soliton-like 
wave-packets can propagate with (quasi)momentum $k$ 
for a long time \cite{malomed96}. 
By means of a variational approach for gaussian 
wavepackets with width $\gamma$ much larger than $1$ 
(the distances are in units of the lattice length), one finds 
that, for $\Lambda > 0$, it is possible to have 
solitonic solutions of the variational equations of motion 
with constant width $\gamma$ if $\cos{k} < 0$ and $\Lambda$ equals 
the critical value \cite{trombettoni01} 
\begin{equation}
\label{lambda_sol}
\Lambda_{sol} \approx  2 \sqrt{\pi} \frac{\mid \cos{k} \mid} 
{\gamma}.
\end{equation}
Numerical simulations confirm that the stability of these wave packet 
is robust for long times.

The generalization of Eq.(\ref{DNLS-retta}) on an arbitrary 
graph is 
\begin{equation} 
\label{DNLS-gen} 
i \frac{\partial \psi_i}{\partial t} = - \frac{1}{2} \sum_{j} A_{i,j}
\psi_{j}+ \Lambda \mid \psi_i \mid ^2 \psi_i :
\end{equation}   
in Eq.(\ref{DNLS-gen}) $A_{i,j}$ is the so-called adjacency matrix 
of the graph \cite{harary69}, which is defined to be $1$ if $i$ and $j$ 
are nearest-neighbours sites, and $0$ otherwise. We shall limit ourself 
to networks obtained inserting simple graphs at two sites of the 
unbranched chain (Fig.1). The two sites of the unbranched 
chain at which the graph is inserted 
are defined to be $n=0$ and $n=\bar{n}$. 
We assume that the soliton is traveling from the left with constant 
velocity $v$, related to $k$ by 
$v \simeq \sin{k}$. Eq.(\ref{DNLS-gen}) is numerically solved with 
$\Lambda=\Lambda_{sol}$ using as initial condition 
\begin{equation}
\psi_n(t=0)={\cal K} e^{-(n-\xi_0)^2/\gamma^2+ik(n-\xi_0)}
\label{in-cond} 
\end{equation}
at the sites $n=\cdots,-1,0,1\cdots$ of the unbranched chain and 
$\psi_i(t=0)=0$ at the sites of the added graph. In Eq.(\ref{in-cond}) 
${\cal K}$ is a normalization factor and 
$\xi_0$ is the initial position of the 
soliton center: we choose $\xi_0<0$ with $\mid \xi_0 \mid \gg 1$ 
and $\pi/2 < k < \pi$, 
so that $v>0$ and the soliton moves 
from the left to the right of the unbranched 
chain. From the numerical solution at very large times (well after 
the collision with the inserted graph) 
the reflection and transmission coefficients $R$
and $T$ are computed by the relations 
$R=\sum_{n<0} \mid \psi_n\mid^2$ and 
$T=\sum_{n>\bar{n}} \mid \psi_n \mid^2$. In the following we shall 
present numerical results for the coefficients $R$ and $T$ obtained 
(for an initial width $\gamma=40$) 
for different values of $k$ and for the networks 
plotted in Fig.1. 

When the soliton is large ($\gamma \gg 1$) 
and fast enough that the soliton-defect collision time is much shorter 
than the soliton dispersion time (i.e. the time scale in which 
the wavepacket will spread in absence of interaction), thus 
one may resort 
to a linear approximation to compute the transmission coefficients 
\cite{cao95,miroshnichenko03} since, in these limits, the soliton 
may be considered as a set of non interacting plane waves  
experiencing scattering on the graph. The 
soliton transmission coefficients 
may be then estimated by computing in the linear regime 
the transmission coefficients of a plane wave across this 
topological defect inserted at two sites. 
Afterwords we shall compare our analytical 
findings with the results coming from the numerical solution 
of Eq.(\ref{DNLS-gen}). 

\section{Transmission coefficients}

\subsection{Loops}

Let consider the situation in which a loop with length $L$ is inserted 
at the two neighbouring sites $n=0$ and $\bar{n}=1$ of the unbranched 
chain [see Figs.1(a,b)]. In the linear approximation, which is expected 
to be reasonable when $L \lesssim \gamma$, the 
transmission coefficients may be determined by considering a plane wave 
solution having 
\begin{equation}
\psi_n=a \, e^{ikn}+b \, e^{-ikn}
\label{left} 
\end{equation}
for $n \le 0$, while for $n \ge 1$ one puts 
\begin{equation}
\psi_n=c \, e^{ikn}.
\label{right} 
\end{equation}
At the sites $\alpha=1,\cdots,L$ of the loop the wavefunction 
is given by
\begin{equation}
\psi_\alpha=d \, e^{ikn}+f \, e^{-ikn}.
\label{loop} 
\end{equation}
The eigenvalue equation to solve is 
\begin{equation}
-\frac{1}{2} \sum_j A_{i,j} \psi_j = {\cal E} \psi_i:
\label{eig} 
\end{equation}
where $i$ and $j$ run on all the sites of the whole network. 
Of course, from Eqs.(\ref{left})-(\ref{right}) one obtains
 ${\cal E}=-\cos{k}$. 
From the continuity in $n=0$, 
$n=1$, $\alpha=1$ and $\alpha=L$ one gets, respectively 
\begin{equation}
-\frac{1}{2} \left( a \, e^{-ik}+ b \, e^{ik}+ c \, e^{ik} + 
d \, e^{ik}+ f \, e^{-ik} \right)=
{\cal E}(a+b)
\label{cont-loop1} 
\end{equation}
\begin{equation}
-\frac{1}{2} \left( a + b + c \, e^{2ik}+ d \, e^{ikL}+ 
f \, e^{-ikL} \right)=
{\cal E} \, c \, e^{ik}
\label{cont-loop2} 
\end{equation}
\begin{equation}
a+b=d+e
\label{cont-loop3} 
\end{equation}
\begin{equation}
c \, e^{ik}=d \, e^{ik(L+1)}+f \, e^{-ik(L+1)}
\label{cont-loop4} 
\end{equation}
From Eqs.(\ref{cont-loop1})-(\ref{cont-loop4}) 
the reflection coefficient 
$R=\mid b/a \mid^2$ is given by:
\begin{equation}
R=\left\vert 
\frac{1+e^{2ikL}-e^{2ik(L+1)}-2e^{ik(L+2)}}  
{1-3e^{2ik}+e^{4ik}-2e^{ik(L+2)}+
e^{2ik(L+2)}+2e^{ik(L+4)}} 
\right\vert^2,
\label{R-loop-L} 
\end{equation}
i.e., $R=R_N/R_D$, where $R_N=5-4\cos(2k)+4 \cos(kL)-
\cos(2k(L+1))-4\cos(k(L+2))$ and 
$R_D=10 [1 -\cos(2k)+\cos(kL) - \cos(k(L+2))] + 
\cos(4k)- 2\cos(k(L-2)) + \cos(2kL) - 3 \cos(2k(L+1))+
\cos(k(L+2))+2 \cos(k(L+4))$.
Of course, if one inserts 
a loop at a single site of the unbranched chain 
(i.e., $n=\bar{n}$), one has to require 
$\psi_{\alpha=1}=\psi_{\alpha=L}$ and 
Eqs.(\ref{cont-loop3})-(\ref{cont-loop4}) simply become 
$a+b=c=d+f$. 

The results obtained for the reflection coefficient 
$R$ from the numerical solution of the DNLSE 
(\ref{DNLS-gen}) and from the linear approximation, 
Eq.(\ref{R-loop-L}), are reported in Fig.2 for $L=2$ (circles) 
and $L=4$ (crosses) and are in remarkable agreement. We see 
that for $L=4$ one has two values of the momenta 
for which there is perfect reflection ($k \approx 2.10$ and 
$k \approx 2.25$): this shows that one can control the soliton 
propagation by properly choosing the topology of the network. 
The average position $\langle n \rangle=\sum_n n \mid 
\psi_n(t) \mid^2$ is plotted vs. time in Fig.3. We see that before 
and after the collision the soliton move with constant velocity: 
with $v=d\langle n \rangle/dt$, one has $v \simeq \sin(k)$ before 
the collision, and $v \simeq - \sin(k)$ after the collision which 
(almost) totally reflects the soliton. The splitting of the soliton 
in transmitted and reflected parts is illustrated in Figs.4-5, 
where we consider
a loop with length $L=2$ and $k=1.8$. In Fig.4 we plot 
$\mid \psi_n \mid^2$ at five different times, including a time ($t=200$) 
in which the soliton hits the loop. In Fig.5 we plot the time evolution 
of the number $N_l$ of particles in the left 
($N_l=\sum_{n \le 0} \mid \psi_n \mid^2$), 
the number $N_r$ of particles in the right 
($N_r=\sum_{n \ge 1} \mid \psi_n \mid^2$) and 
the number $N_{loop}$ of particles in the loop 
($N_{loop}=\sum_{\alpha=1}^{L} \mid \psi_{\alpha} \mid^2$): one sees that 
for $t \approx 200$ the number of particles in the loop increases 
and after the reflection decreases.

In the simplest case ($L=1$) a single 
extra site is attached to the sites $n=0$ and $\bar{n}=1$ of the 
unbranched chain, as in Fig.1(a): 
then Eq.(\ref{R-loop-L}) simplifies to
\begin{equation}
R=\frac{2 \cos^2{(k/2)}}
{2+\cos{k}-\cos{(3k)}}.
\label{R-loop-1} 
\end{equation}
The comparison between the results for $R$ 
from the numerical solution of the DNLSE 
(\ref{DNLS-gen}) and from 
Eq.(\ref{R-loop-1}) is reported in Fig.6, showing also in this case 
a good agreement,

We observe that the extra sites of the inserted graph can 
be view as external Fano degrees of freedom coupled 
to the chain \cite{miroshnichenko03,flach03,miroshnichenko05}. 
In particular, in \cite{miroshnichenko05} an additional 
discrete state is coupled to the sites of a straight linear chain: 
this would correspond in our description to a site 
linked to all the sites of the unbranched chain.

\subsection{Bubbles}

We refer in this Section to inserted $p$-bubble graphs, i.e. 
to loops with length $L=1$ inserted $p$-times at two sites 
of the unbranched chains which are distant $2$ [see Fig.1(c)]. 
To fix the notations the sites of the unbranched chain are 
defined to be $n=\cdots,-1,0_1,1,\cdots$, and other 
$p-1$ sites $0_2,0_3,\cdots,0_p$ are linked to the sites 
$-1$ and $1$. Of course, if $p=1$, we have the simple chain. 

The coefficient $R$ may be obtained in the linear approximation 
in the following way: we assume 
$\psi_n=a \, e^{ikn}+b \, e^{-ikn}$ for $n \le -1$,
$\psi_n=c \, e^{ikn}$ for $n \ge 1$, and 
$\psi_0 \equiv \psi_{0_1}= \cdots = \psi_{0_p}$ 
(for symmetry, all the sites inside the bubble are equivalent). 
The eigenvalue equation (\ref{eig}), with 
${\cal E}=-\cos{k}$, in the site $-1$ reads 
$-(1/2)(\psi_{-2}+p\psi_0)={\cal E} \psi_{-1}$, i.e., 
\begin{equation}
-\frac{1}{2} \left( a \, e^{-2ik}+ b \, e^{2ik}+ p \, \psi_0 \right)=
{\cal E} (a \, e^{-ik}+b \, e^{ik})
\label{cont-bubble1}.  
\end{equation}
Similarly, Eq.(\ref{eig}) in the site $+1$ gives
\begin{equation}
-\frac{1}{2} \left( p \, \psi_0 + c \, e^{2ik} \right)=
{\cal E} \, c \, e^{-ik}
\label{cont-bubble2}.  
\end{equation}
In a site inside the bubble, Eq.(\ref{eig}) reads
\begin{equation}
-\frac{1}{2} \left( a \, e^{-ik} +  b \, e^{ik} + c \, e^{-ik} 
\right)=
{\cal E} \psi_0
\label{cont-bubble3}.  
\end{equation}
From Eqs.(\ref{cont-bubble1})-(\ref{cont-bubble3}) one gets 
\begin{equation}
R=\frac{1}{1+\left( \frac{p}{p-1}\right)^2 \tan^2{k}}:
\label{R-bubble} 
\end{equation}
of course, when $p=1$, no reflection occurs. Eq.(\ref{R-bubble}) 
shows that increasing $k$ from $\pi/2$ to $\pi$ (i.e., decreasing 
the velocity), the reflection increases: slower solitons are more 
reflected. In this meaning, the $p$-bubble is an high-pass, i.e. 
only solitons with high velocity are transmitted: by varying $p$ one can 
control the width of the range of transmitted velocities. In Fig.7 
we plot the reflection coefficient $R$ for $p=2$ and $p=20$, showing 
that a larger $p$ make smaller the range of transmitted velocities. 
As in Figs.2 and 6, solid lines correspond to the analytical estimate
[given by Eq.(\ref{R-bubble})], and numerical results from the DNLSE 
(\ref{DNLS-gen}) are expressed by 
circles ($p=2$) and crosses ($p=20$).  

\subsection{Separate links}

In this Section we consider the effect on the soliton propagation 
of two extra sites linked to two sites the unbranched chain, as in 
Fig.1(d). To fix notations, we suppose that the two extra sites, 
$\alpha$ and $\beta$, are linked respectively to $n=0$ and 
$\bar{n}=L$. 

To obtain the coefficient $R$ in the linear approximation, one proceeds 
as before: one assumes $\psi_n=a \, e^{ikn}+b \, e^{-ikn}$ for 
$n \le 0$,
$\psi_n=c \, e^{ikn}$ for $n \ge L$, and 
$\psi_n= d \, e^{ikn} + f \, e^{ikn}$ for $n=1,\cdots,L-1$. 
The eigenvalue equation (\ref{eig}), with   
${\cal E}=-\cos{k}$, in the sites $0$, $L$, $A$, $B$, $1$ and $L+1$ 
reads respectively
\begin{equation}
-\frac{1}{2} \left( a \, e^{-ik}+ b \, e^{ik}+ \psi_\alpha  + 
d \, e^{ik} + f \, e^{-ik} \right)=
{\cal E} \,  (a + b)
\label{cont-sep1}
\end{equation}
\begin{equation}
-\frac{1}{2} \left( d \, e^{ik(L-1)}+ f \, e^{-ik(L-1)}+ \psi_\beta  + 
c \, e^{ik(L+1)} \right)=
{\cal E} \, c \, e^{ikL}
\label{cont-sep2}
\end{equation}
\begin{equation}
-\frac{1}{2} \left( a + b \right)=
{\cal E} \, \psi_\alpha
\label{cont-sep3}
\end{equation}
\begin{equation}
-\frac{1}{2} \, c \, e^{ikL} =
{\cal E} \, \psi_\beta
\label{cont-sep4}
\end{equation}
\begin{equation}
a+b=d+e
\label{cont-sep5}
\end{equation}
\begin{equation}
c \, e^{ikL}=d \, e^{ikL} + f \, e^{-ikL}
\label{cont-sep6}.
\end{equation}
One has seven unknowns ($a$, $b$, $c$, $d$, $f$, $\psi_\alpha$ and 
$\psi_\beta$) 
and the six equations (\ref{cont-sep1})-(\ref{cont-sep6}) 
(the remaining is provided by the normalization). Solving for $b/a$ 
one gets
\begin{equation}
R=\left\vert 
\frac{1-e^{2ik}-e^{4ik}+e^{2ikL}+e^{2ik(L+1)}-e^{2ik(L+2)}}
{e^{2ik(L+2)}-\left( e^{2ik}+e^{4ik}-1\right)^2} 
\right\vert^2.
\label{R-sep-L}
\end{equation}
for $L=1$ Eq.(\ref{R-sep-L}) simplifies to 
\begin{equation}
R=\frac{\left( 1+ 2 \cos{(2k)} \right)^2 }
{9\cos^2{k}+\left( \sin{k}+2 \sin{(3k)}\right)^2} 
\label{R-sep-1}
\end{equation}
and for $L=2$ to 
\begin{equation}
R=\frac{\left( 1- 2 \cos{(2k)} \right)^2 }
{4\sin^2{(2k)}+\left( 2\cos{(2k)}-1\right)^2}. 
\label{R-sep-2}
\end{equation}
Notice that if you have only a site linked to a site 
of the unbranched chain, one obtains in the linear limit 
\cite{miroshnichenko03,chaos}
\begin{equation}
R=\frac{1}{1+4 \sin^2{(2k)}}:
\label{R-single}
\end{equation}
a comparison of Eq.(\ref{R-single}) with Eq.(\ref{R-sep-L}) 
shows the remarkable effect (also for large $L$) 
of the second added link on the transmission properties. In Fig.8 
we plot the reflection coefficient $R$ for $L=1$ and $L=2$, showing 
that in both situations one has a peak in the transmission 
($R \approx 0$) for two different values: the two separate links 
behave approximately as transmission filters, i.e., allowing 
for the transmission of solitons with certain velocities. 
Increasing the distance $L$ between the sites at which the added sites 
are linked the number of such transmission peaks in turn increases. 
In Fig.8 the solid lines correspond to the analytical estimate
[given by Eqs.(\ref{R-sep-1})-(\ref{R-sep-2})], 
and numerical results from the DNLSE 
(\ref{DNLS-gen}) are expressed by 
circles ($L=1$) and crosses ($L=2$).  

\section{Conclusions}

In this paper we studied the propagation of large-fast solitons 
on complex chains built by inserting finite graphs at two sites 
of an unbranched chain. The inserted graph 
is seen by the soliton like an extended topological defect: 
the transmission properties of the solitons strongly depend 
on the structure of the inserted graph. We considered simple 
inserted graphs: loops, bubbles and single links. For inserted loops 
peaks for perfect reflection occurs, while 
for bubbles the inserted graph behaves as a high-pass filter. 
When two added sites are linked to two sites having distance 
$L$ between them, transmission peaks appears and 
increasing the $L$ the number of such transmission peaks increases. 
In all these situations we compared numerical findings with 
the results of an analytical linear approximation, obtaining 
a good agreement.

In conclusion we think that the study of nonlinear dynamical systems 
on complex networks is a wide subject to investigate, and that the main 
motivation of such study is that the network topology provides 
a natural tool to control the nonlinear dynamics of wavepackets. 
In particular, we mention as possible interesting future studies the 
study of the propagation of very localized breathers on complex 
networks and the nonlinear trapping of solitonic solution in chains 
with topological defects.

{\em Acknowledgments:} We thank P. G. Kevrekidis and 
B. A. Malomed for stimulating discussions.

\begin{figure}[h]
\centerline{\psfig{
figure=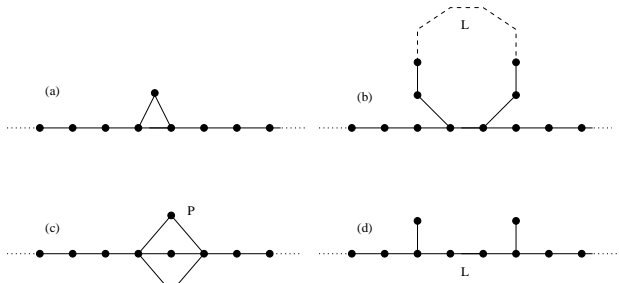,width=82mm,angle=0}}
\caption{Inserting simple graphs at two sites of a linear chain: 
(a) a single extra site attached to two sites 
of the chain; (b) a loop with length $L$; 
(c) a bubble with $p=3$ sites; 
(d) two single links attached at two sites with distance $L=3$.}
\end{figure}

\begin{figure}[h]
\centerline{\psfig{
figure=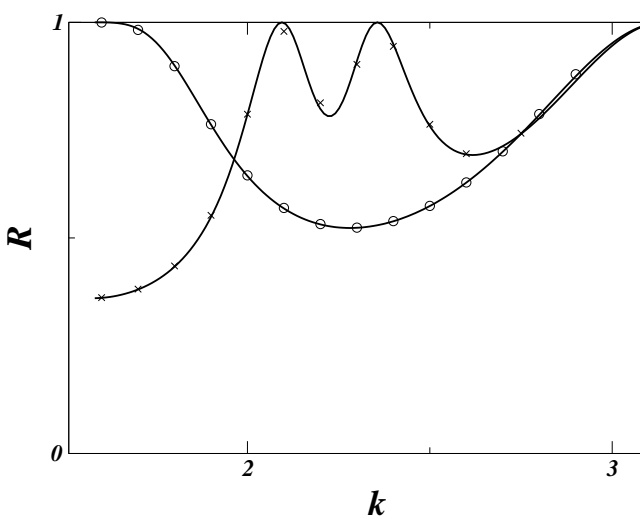,width=82mm,angle=270}}
\caption{Reflection coefficient $R$ as a function of $k$ 
(with $k$ between $\pi/2$ and $\pi$) when loops with length $2$ 
and $4$ are attached. 
Empty circles ($L=2$) and crosses ($L=4$) 
correspond to the numerical solution of Eq.(\ref{DNLS-gen}): 
as initial condition we choose a Gaussian with initial 
width $\gamma=40$ and momentum $k$ (see text). 
Solid lines correspond to the analytical prediction 
(\ref{R-loop-L}).}
\end{figure}

\begin{figure}[h]
\centerline{\psfig{
figure=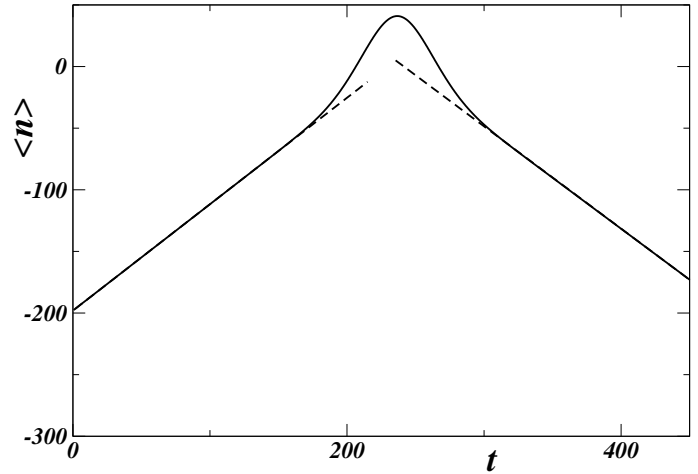,width=82mm,angle=270}}
\caption{Center of mass position $\langle n \rangle$ vs. time 
for $k=2.1$ and a loop having $L=4$, which corresponds to perfect 
reflection. The solid line correspond to the numerical solution 
of Eq.(\ref{DNLS-gen}), and the dashed lines to free motion of the soliton 
with absolute value of the velocity $\mid v \mid=\sin{k}$.}
\end{figure}

\begin{figure}[h]
\centerline{\psfig{
figure=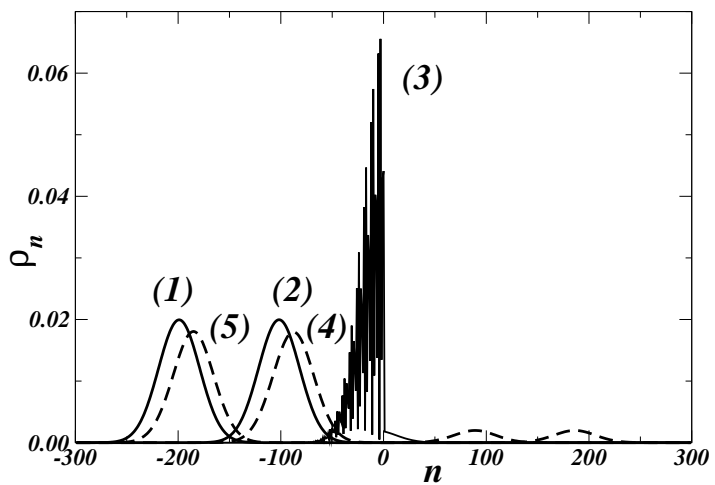,width=82mm,angle=270}}
\caption{Soliton propagation obtained from Eq.(\ref{DNLS-gen}) 
[for momentum $k_=1.8$ and width $\gamma_0=40$] 
through an inserted loop of length $2$. 
The soliton profile is plotted for 
$z=0,100,200,300,400$ corresponding to $(1)\cdots (5)$.}
\end{figure}

\begin{figure}[h]
\centerline{\psfig{
figure=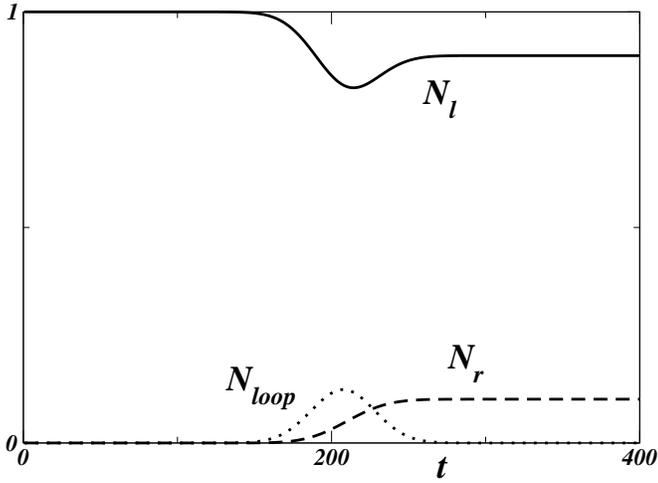,width=82mm,angle=270}}
\caption{Time evolution from the numerical solution of 
Eq.(\ref{DNLS-gen}) for the number of particles 
on the left of the loop ($N_l$), on the right of the loop 
($N_r$) and on the loop 
($N_{loop}$) for $L=2$ and $k=1.8$. 
Around $t=200$ particles enter the loop 
(compare with the previous figure).}
\end{figure}

\begin{figure}[h]
\centerline{\psfig{
figure=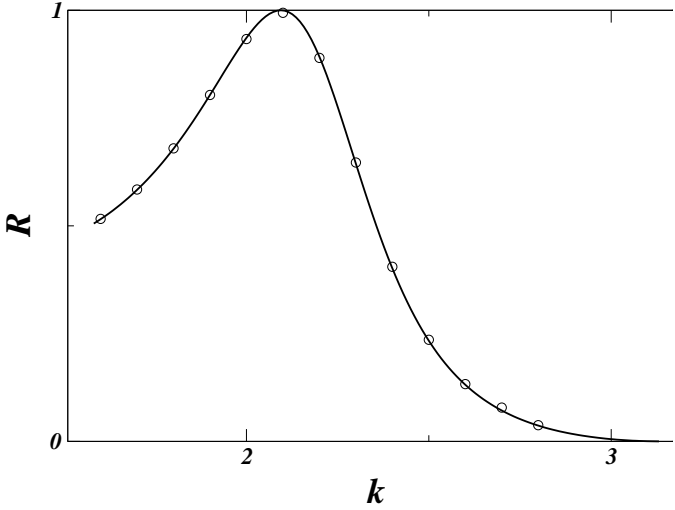,width=82mm,angle=270}}
\caption{Reflection coefficient $R$ as a function of $k$ 
(with $k$ between $\pi/2$ and $\pi$) for a loop 
made of a single site ($L=1$). 
Empty circles correspond to the numerical solution of Eq.(\ref{DNLS-gen}) 
and the solid lines correspond to the analytical prediction 
(\ref{R-loop-1}).}
\end{figure}

\begin{figure}[h]
\centerline{\psfig{
figure=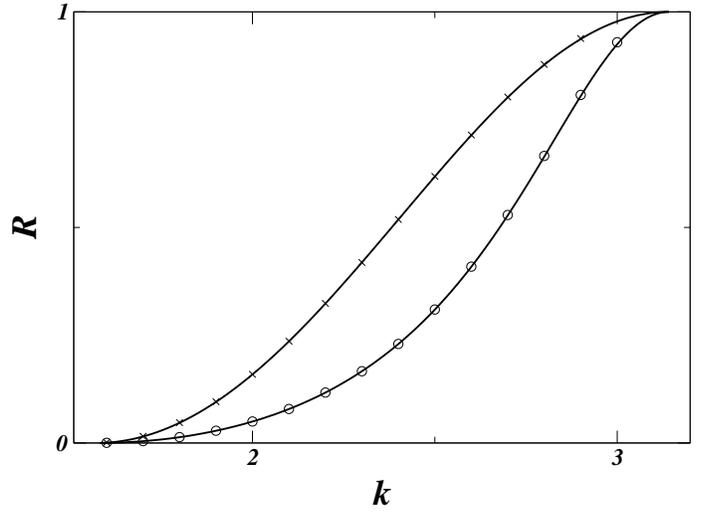,width=82mm,angle=270}}
\caption{Reflection coefficient $R$ as a function of $k$ 
when $p$-bubbles with $p=2$ and $p=20$ are attached. 
Empty circles ($p=2$) and crosses ($p=20$) 
correspond to the numerical solution of Eq.(\ref{DNLS-gen}), 
solid lines correspond to the analytical prediction 
(\ref{R-bubble}).}
\end{figure}

\begin{figure}[h]
\centerline{\psfig{
figure=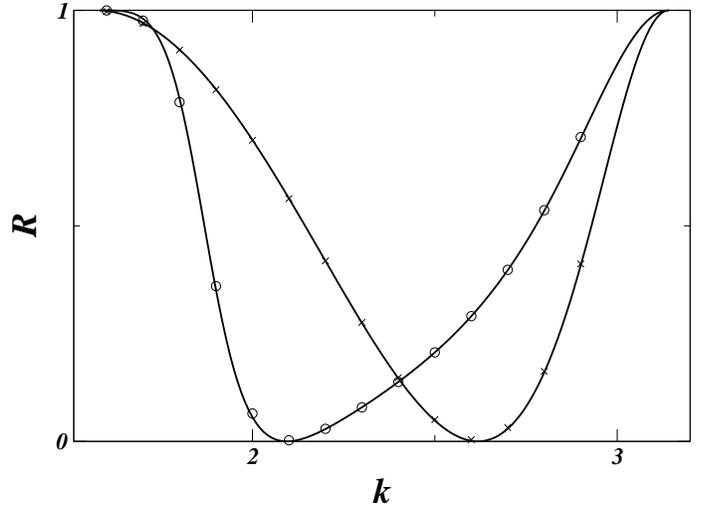,width=82mm,angle=270}}
\caption{Reflection coefficient $R$ as a function of $k$ 
when separate sites are linked to sites 
distant $L=1$ and $L=2$. Empty circles ($L=1$) and crosses ($L=2$) 
correspond to the numerical solution of Eq.(\ref{DNLS-gen}), 
solid lines correspond to the analytical predictions 
(\ref{R-sep-1})-(\ref{R-sep-2}).}
\end{figure}

\end{multicols}

\end{document}